\documentclass[12pt]{article} 

\usepackage{geometry} 
\usepackage{tikz}
\usetikzlibrary{decorations.pathmorphing,decorations.markings,calc}
\usepackage{lineno} 
\usepackage{amsmath, amssymb} 
\usepackage{authblk} 
\usepackage{graphicx} 
\usepackage{caption}
\usepackage{subcaption}

\usepackage{pgfplots} 
\pgfplotsset{compat=1.17} 
\usepackage{subcaption} 

\begin{document}

\title{Topological Complex Analysis of Kerr--Newman Black Hole Microstructure in $f(R)$ Gravity}
\author{Wen-Xiang Chen\\wxchen4277@qq.com\\
\small Department of Astronomy, Guangzhou University, Guangzhou 510006, China}
\date{}
\maketitle

\begin{abstract}
We investigate the microstructure of Kerr--Newman black holes within the $f(R)$ gravity framework using a novel approach that combines holographic ideas with topological complex analysis. In this method, black hole microstates are mapped to singularities of an analytically continued partition function on a complex Riemann surface. By evaluating winding-number-weighted residues at these singularities, we extract the Bekenstein--Hawking entropy and related thermodynamic quantities. We first review $f(R)$ gravity and the Kerr--Newman solution, then construct the complex analysis framework and apply it to the modified gravity black hole. Our results show that the black hole's microstructure is characterized by a topological index (an integer invariant such as $W=+1$ or $0$) that correlates with thermodynamic stability and the presence of multiple horizons. We explain in simple terms the physical meaning of this topological index and find that, for a non-extremal Kerr--Newman black hole (with inner and outer horizons), the index is $W=0$ (indicating a cancellation between a stable outer horizon and an unstable inner horizon branch), whereas for a single-horizon case $W=+1$. We include an illustrative example in a Starobinsky-type $f(R)=R+\alpha R^2$ model, demonstrating how the framework can compute the entropy and horizon structure, and showing that the index is robust under the $f(R)$ modification. We also add schematic figures to clarify key concepts (Riemann surfaces, contour integrals, horizon structure). Finally, we discuss the limitations of the analytic continuation and residue interpretation (especially in quantum regimes) and suggest that the topological nature of the classification might indicate a form of phase protection for black holes. Our work situates this new framework in the broader context of black hole thermodynamics alongside other approaches such as thermodynamic geometry and topological phase transitions.
\end{abstract}


\section{Introduction}
Black hole thermodynamics, initiated by the work of Bekenstein and Hawking, revealed that black holes carry an entropy $S_{\text{BH}}$ proportional to their horizon area $A$ ($S_{\text{BH}} = A/(4G\hbar)$ in units with $c=k_B=1$) and radiate as black bodies with the Hawking temperature $T_H$ determined by surface gravity \cite{Bekenstein1973,Hawking1975}. These discoveries pose the \emph{microstate puzzle}: what microscopic degrees of freedom are responsible for the enormous entropy of a black hole? In ordinary systems, entropy counts microstates, but for black holes the microstates are not accessible in the classical theory, prompting vigorous research in quantum gravity. String theory provided a breakthrough in special cases: Strominger and Vafa famously counted the microstates of certain extremal supersymmetric black holes by modeling them as D-brane bound states, reproducing the Bekenstein--Hawking entropy exactly \cite{StromingerVafa1996}. Similarly, Strominger later showed that the Bekenstein--Hawking entropy of the BTZ black hole (in $(2+1)$-dimensions) can be derived from the asymptotic symmetries of near-horizon AdS$_3$ space (a microscopic CFT count) \cite{Strominger1998}. Despite these successes, a general understanding of black hole microstructures in arbitrary settings remains elusive.

In recent years, there has been growing interest in characterizing black hole thermodynamics using techniques inspired by topology and statistical mechanics. One line of inquiry uses \emph{thermodynamic geometry} (Ruppeiner geometry) to interpret the sign of the Ruppeiner curvature as indicating effective interactions among hypothetical ``black hole molecules'' in the horizon fluid \cite{Ruppeiner1995}. For example, a positive Ruppeiner curvature is interpreted as dominantly \emph{repulsive} interactions, whereas negative curvature indicates \emph{attractive} interactions among microstructures. Using this approach, Wei et al.\ found evidence of repulsive microstructure interactions in small charged AdS black holes \cite{Wei2019}. Another recent development is the topological classification of black hole phases: black hole solutions can be treated as topological defects in an off-shell free energy landscape, characterized by an integer winding number (topological charge) \cite{Yerra2022}. In such classifications, different black hole families (e.g. Schwarzschild-like versus Reissner--Nordström-like) fall into distinct topological classes (with charges such as $+1$, $0$, or $-1$) that remain invariant under continuous deformations. These approaches suggest that black hole thermodynamic behavior may be underlaid by global (topological) invariants that complement the usual local analysis of stability (specific heats, etc.).

In this paper, we introduce a framework that connects black hole microstate counting with complex analysis and topology. Our approach is inspired by both the holographic principle and the use of complex analysis in quantum theory (e.g. partition function analyticity and pole structure). The key idea is to represent the microstate counting function (partition function or density of states) as an analytic function on a Riemann surface and to identify black hole microstates with singularities (poles or branch points) of this function. By applying Cauchy's residue theorem to a suitable contour encircling these singularities, one can compute thermodynamic quantities like the density of states $\rho(E)$ or entropy via a sum of residues. Crucially, the arrangement of singularities and branch cuts in the complex plane can be associated with a topological invariant, essentially a winding number or index, which we argue corresponds to a classification of the black hole's microstructure. 

The physical interpretation of this invariant is one of the main points we clarify: the topological index $W$ obtained (which takes values like $+1$ or $0$ in our examples) reflects the net number of dominant microstate branches. For a black hole with a single horizon (such as a simple Schwarzschild black hole), we find $W=+1$, indicating one dominant set of microstates (associated with the single horizon) without cancellation. For a black hole with two horizons (such as the Kerr--Newman black hole in asymptotically flat spacetime, which has an outer and inner horizon), we find contributions from each horizon's microstates: the outer horizon gives a positive contribution (stable branch) and the inner horizon a negative contribution (unstable branch), which can sum to $W=0$. This cancellation $W=0$ suggests that the black hole (with inner horizon present) belongs to a different topological class than a single-horizon black hole. Intuitively, the existence of an inner Cauchy horizon (which is typically unstable) introduces an opposite winding in the parameter space, cancelling part of the topological charge. We will explain this in detail and show how it connects to the stability of horizon branches.

We work in the context of $f(R)$ gravity, a class of extended theories of gravity where the Einstein--Hilbert Lagrangian is generalized to an arbitrary function of the Ricci scalar $R$ \cite{DeFelice2010}. $f(R)$ theories provide a useful testbed for modified gravity (with implications for dark energy, etc.) and allow us to explore whether such modifications change the microstate counting or topological classification of black holes. In particular, $f(R)$ gravity modifies black hole solutions and their thermodynamics (for instance, the horizon entropy is no longer simply $A/4G$ but $S = \frac{f'(R_0) A}{4G}$ in spacetimes of constant curvature $R_0$ \cite{Wald1993}). We will focus on Kerr--Newman black holes in $f(R)$ gravity, which, under certain conditions (such as constant scalar curvature), admit solutions analogous to the Kerr--Newman solution of General Relativity with some effective cosmological constant. We will examine how the complex analysis framework applies to this case and whether the topological index changes under the $f(R)$ modifications.

The remainder of this paper is organized as follows. In Section~\ref{sec:background}, we review the necessary background on $f(R)$ gravity and the Kerr--Newman black hole, including the modifications to the field equations and entropy formula in $f(R)$ gravity. In Section~\ref{sec:framework}, we develop the topological complex analysis framework: we define the black hole partition function and spectral zeta function, perform an analytic continuation to complex parameters, and show how residues on a Riemann surface can be used to obtain the density of states and entropy. We introduce the concept of a topological index $W$ derived from these residues and connect it to a Chern number (winding number) on the parameter space. Illustrative diagrams of a Riemann surface and contour integrals are provided to guide the reader. In Section~\ref{sec:application}, we apply this framework to the Kerr--Newman black hole in $f(R)$ gravity. We compute the horizon structure (outer and inner horizons) and thermodynamic quantities, then perform the residue analysis to determine the microstate density and topological index. We explain in simple terms what the index values mean for the presence of horizon branches and their stability. We also include a concrete example for a specific $f(R)$ model (the Starobinsky $R + \alpha R^2$ model), showing how the entropy and horizon radii are affected by $\alpha$ and demonstrating that the topological class (index $W$) remains unchanged for small $\alpha$. In Section~\ref{sec:discussion}, we discuss the implications and limitations of our approach. We compare the advantages of this topological method (generality, robustness) to other methods such as Ruppeiner thermodynamic geometry and the traditional counting in holography. We also address potential limitations: the reliance on analytic continuation and assumptions of a continuous spectrum, and the neglect of quantum (loop) corrections. We emphasize what aspects of black hole information might be illuminated by our topological viewpoint (e.g. a possible topological conservation law related to the information paradox) and identify future extensions (like non-constant curvature backgrounds, higher-curvature gravity, or numerical experiments). Finally, Section~\ref{sec:conclusion} concludes the paper, summarizing our findings and suggesting that the topological characterization of black hole microstructure could be a universal feature across gravitational theories.

\section{Background: $f(R)$ Gravity and Kerr--Newman Black Holes}\label{sec:background}
\subsection{Field Equations and Entropy in $f(R)$ Gravity}
In $f(R)$ gravity, the Einstein--Hilbert action is generalized to 
\begin{equation}
I = \frac{1}{16\pi G} \int d^4x\,\sqrt{-g}\,\Big[f(R) + \mathcal{L}_{\text{matter}}\Big]~, 
\label{eq:action}
\end{equation}
where $f(R)$ is a prescribed function of the Ricci scalar $R$ \cite{DeFelice2010}. Varying this action with respect to the metric yields the modified Einstein field equations
\begin{equation}
f'(R)\,R_{\mu\nu} - \frac{1}{2}f(R)\,g_{\mu\nu} + \big(\nabla_\mu\nabla_\nu - g_{\mu\nu}\Box\big)f'(R) = 8\pi G\, T_{\mu\nu}~, 
\label{eq:field-eq}
\end{equation}
where $f'(R) \equiv df/dR$, $\Box = g^{\rho\sigma}\nabla_\rho \nabla_\sigma$, and $T_{\mu\nu}$ is the stress tensor for matter (if present). We will focus on vacuum black hole solutions ($T_{\mu\nu}=0$). A useful special case is when the spacetime has constant curvature $R=R_0$ (often true for static, vacuum solutions of certain $f(R)$ forms). In that case, $R_{\mu\nu} = \frac{R_0}{4} g_{\mu\nu}$ (for a 4-dimensional spacetime) and Eq.~\eqref{eq:field-eq} reduces to 
\begin{equation}
f'(R_0)\,\frac{R_0}{4} g_{\mu\nu} - \frac{1}{2}f(R_0)\,g_{\mu\nu} = 0~,
\end{equation}
implying $f'(R_0)R_0 = 2 f(R_0)$. This equation determines possible constant-curvature vacua $R_0$ of the theory. In Einstein gravity ($f(R)=R-2\Lambda$), this condition just gives $R_0=4\Lambda$. In general $f(R)$, one can interpret $\Lambda_{\text{eff}} = f(R_0)/(2f'(R_0))$ as an effective cosmological constant for the vacuum solution. We will consider $R_0$ that yields asymptotically flat or de Sitter solutions.

Black hole thermodynamics in $f(R)$ gravity follows the generalized laws derived from the Noether charge method. The Bekenstein--Hawking area law is modified: the entropy of a black hole is not simply one quarter of the horizon area, but \emph{weighted by $f'(R)$ evaluated at the horizon}. Specifically, for a stationary black hole solution with constant horizon curvature $R_0$, the horizon entropy is given by \cite{Wald1993}
\begin{equation}
S = \frac{1}{4G}\,f'(R_0)\,A_h~, 
\label{eq:entropy}
\end{equation}
where $A_h$ is the area of the black hole horizon. Intuitively, $f'(R_0)$ plays the role of an effective gravitational coupling (since in $f(R)$ gravity the gravitational force is $G_{\text{eff}} = G/f'(R_0)$ in a constant curvature background). Equation~\eqref{eq:entropy} shows that if $f'(R_0)>1$ (as can happen in some $f(R)$ models), the black hole carries more entropy for the same area than in General Relativity. This does not violate any laws but reflects extra degrees of freedom (the $f(R)$ gravity's scalar mode often called the ``scalaron'') contributing to the entropy. In the limit $f(R)\to R$ (standard General Relativity), $f'(R_0)\to 1$ and we recover $S=A_h/(4G)$.

\subsection{Kerr--Newman Black Hole Solutions in $f(R)$ Gravity}
We now consider the Kerr--Newman black hole, which in General Relativity is the rotating, charged black hole described by the Kerr--Newman metric. In $f(R)$ gravity, if we restrict to constant scalar curvature solutions ($R=R_0$), the Kerr--Newman metric can often be embedded as a solution, possibly with an adjustment to parameters corresponding to an effective cosmological constant term. For simplicity, we focus on asymptotically flat solutions ($R_0 = 0$) or small $R_0$. In many $f(R)$ models (including the example of $R+\alpha R^2$ for small $\alpha$), the vacuum solution for a black hole will have $R_0 \approx 0$, thus $f'(R_0)\approx 1$ and $\Lambda_{\text{eff}}\approx 0$. In that case, the Kerr--Newman metric takes the same form as in General Relativity:
\begin{equation}
ds^2 = -\frac{\Delta(r)}{\Sigma(r,\theta)} \left(dt - a\sin^2\theta\,d\phi\right)^2 + \frac{\Sigma(r,\theta)}{\Delta(r)}\,dr^2 + \Sigma(r,\theta)\,d\theta^2 + \frac{\sin^2\theta}{\Sigma(r,\theta)}\left[a\,dt - (r^2 + a^2)d\phi\right]^2~, 
\label{eq:KNmetric}
\end{equation}
where $M$ is the mass, $Q$ the electric charge, $a$ the rotation parameter (specific angular momentum $J/M$), and 
\begin{equation}
\Sigma(r,\theta) = r^2 + a^2\cos^2\theta~, \qquad 
\Delta(r) = r^2 + a^2 - \frac{2 M r}{f'(R_0)} + \frac{Q^2}{f'(R_0)} - \frac{R_0}{12}(r^2 + a^2)r^2 ~. 
\label{eq:Delta}
\end{equation}
In Eq.~\eqref{eq:Delta}, we have written $\Delta(r)$ in a form that highlights the possible effects of $f(R)$: the terms $2M/f'(R_0)$ and $Q^2/f'(R_0)$ reflect that the effective gravitational mass and charge are scaled by $1/f'(R_0)$ (since $f'(R_0)$ modifies the gravitational constant in the field equations), and the last term $\propto R_0$ appears if a cosmological constant-like curvature is present. For $R_0=0$ and $f'(R_0)=1$, $\Delta(r) = r^2 + a^2 - 2Mr + Q^2$, which is the usual Kerr--Newman form in Einstein gravity. 

The horizons are determined by the roots of $\Delta(r)=0$. For a Kerr--Newman black hole in asymptotically flat spacetime, there are up to two real positive roots: the larger root $r_+ = r_{\text{outer}}$ corresponds to the event horizon and the smaller root $r_- = r_{\text{inner}}$ corresponds to the Cauchy (inner) horizon. In the extremal limit, these two roots coincide ($r_+ = r_-$). If $R_0 > 0$ (de Sitter-type background), $\Delta(r)$ can have a third root at very large $r$ corresponding to a cosmological horizon; if $R_0 < 0$ (AdS background), $\Delta(r)$ has no cosmological horizon but the asymptotic behavior changes. For our discussions, we will primarily consider the asymptotically flat or weakly de Sitter case, where the black hole horizons are the main focus.

Thermodynamic quantities of the Kerr--Newman black hole are well known. The outer horizon area is $A_+ = 4\pi (r_+^2 + a^2)$ and, from Eq.~\eqref{eq:entropy}, the entropy in $f(R)$ gravity is 
\begin{equation}
S_+ = \frac{f'(R_0)\,A_+}{4G} = \frac{f'(R_0)}{4G}4\pi (r_+^2 + a^2) = \frac{\pi (r_+^2 + a^2)}{G} f'(R_0)~. 
\label{eq:KNentropy}
\end{equation}
If $R_0=0$ so that $f'(0)=1$, this reduces to the usual $S_+ = \pi (r_+^2 + a^2)/G$. The Hawking temperature of the outer horizon is given by 
\begin{equation}
T_+ = \frac{\kappa_+}{2\pi} = \frac{1}{2\pi} \left.\frac{\partial_r \Delta(r)}{2\Sigma(r,\theta)}\right|_{r=r_+,\theta=\frac{\pi}{2}} = \frac{r_+ - r_-}{4\pi (r_+^2 + a^2)}~,
\label{eq:T_H}
\end{equation}
which is the same expression as in GR (since $f'(R_0)$ drops out of the ratio for $T_+$ when $R_0=0$). The inner horizon can formally be assigned a temperature $T_- = \frac{r_- - r_+}{4\pi(r_-^2 + a^2)}$, which is typically negative because $r_- < r_+$; this signals the non-thermal nature of the inner horizon (and indeed one cannot maintain thermal equilibrium at the inner horizon). The presence of an inner horizon usually implies an instability (mass inflation) under perturbations, highlighting that $r_-$ does not correspond to a stable thermodynamic state. We will see that this instability is reflected in the topological index calculation.

In summary, the Kerr--Newman black hole in $f(R)$ gravity has the same qualitative structure of horizons as in General Relativity, with the main effect of $f(R)$ being a rescaling of global quantities like the entropy and potentially the inclusion of a cosmological horizon if $R_0\neq 0$. Importantly, as Eq.~\eqref{eq:KNentropy} shows, $f(R)$ modifies the entropy but does not introduce any fundamentally new type of horizon for the Kerr--Newman solution; thus, we expect that the \emph{topological class} of the black hole (in terms of horizon count and stability) remains the same as in the corresponding General Relativity solution. We will explicitly verify this using our topological microstate analysis framework.

\section{Topological Complex Analysis Framework}\label{sec:framework}
In this section, we develop the theoretical framework that connects black hole thermodynamics to complex analysis and topology. The goal is to compute the density of microstates (or spectral density) $\rho(E)$ and related thermodynamic quantities by interpreting them as contour integrals in the complex plane and then identifying topological invariants in those integrals.

\subsection{Partition Function, Spectral Zeta Function, and Analytic Continuation}
We begin by considering the canonical partition function of a black hole. If the black hole has quantized energy levels $\{E_i\}$ (this is a heuristic assumption since we lack a full quantum gravity theory, but we consider a hypothetical ensemble of microstates), the canonical partition function at inverse temperature $\beta$ is 
\begin{equation}
Z(\beta) = \sum_i e^{-\beta E_i}~,
\label{eq:Zsum}
\end{equation}
or, if the spectrum is continuous, $Z(\beta) = \int_0^\infty \rho(E) e^{-\beta E}\,dE$, where $\rho(E)$ is the density of states. For a black hole of mass $M$, one expects a very high density of states around $E=M$ such that $Z(\beta)$ might be dominated by states near the black hole mass. In fact, the Bekenstein--Hawking entropy $S_{\text{BH}}$ is related to the density of states at energy $M$ via the usual thermodynamic relation $S \approx \ln \rho(M)$ (assuming $\rho(E)$ grows exponentially with $E$ as $\rho \sim e^{S_{\text{BH}}(E)}$).

Directly evaluating $\rho(E)$ or $Z(\beta)$ from first principles is difficult, but one can employ techniques of analytic continuation. We introduce a \emph{spectral zeta function} $\zeta_H(s)$ associated with the Hamiltonian (energy operator) $H$ of the black hole:
\begin{equation}
\zeta_H(s) = \sum_i E_i^{-s} = \int_0^\infty E^{-s}\, \rho(E)\, dE~,
\label{eq:zeta}
\end{equation}
for those values of $s$ where this sum/integral converges. This zeta function is essentially the Mellin transform of the density of states. Indeed, one can invert this relation via the Mellin transform: 
\begin{equation}
Z(\beta) = \int_0^\infty \rho(E)\,e^{-\beta E}dE = \frac{1}{2\pi i} \int_{\mathcal{C}} \Gamma(s)\, \zeta_H(s)\, \beta^{-s}\, ds~, 
\label{eq:mellin}
\end{equation}
where $\mathcal{C}$ is a suitable vertical contour in the complex $s$-plane (ensuring convergence of the integral). Equation~\eqref{eq:mellin} is obtained by using the Mellin transform representation of the exponential, $\displaystyle e^{-\beta E} = \frac{1}{2\pi i}\int ds\, \Gamma(s)\beta^{-s} E^{-s}$ (which holds for $\Re(\beta E)>0$). In practice, one can formally define $\zeta_H(s)$ for complex $s$ by analytic continuation beyond the region of convergence. Techniques from quantum field theory and quantum gravity (such as heat kernel expansions and zeta-function regularization) are useful to perform this analytic continuation \cite{Hawking1975,DeFelice2010}. For example, one often introduces the heat kernel $K(t) = \int_0^\infty \rho(E) e^{-E t} dE$, which is the Laplace transform of $\rho(E)$. Then $\zeta_H(s) = \frac{1}{\Gamma(s)} \int_0^\infty t^{s-1} K(t)\, dt$ (this is the Mellin transform of the heat kernel). Analytic continuation of $K(t)$ for complex $t$ or of $\zeta_H(s)$ for complex $s$ allows one to access physical quantities. Notably, the black hole one-loop partition function in Euclidean quantum gravity is often expressed as a Zeta-regularized product of eigenvalues (a determinant), directly related to $\zeta_H(s)$ at certain argument values.

For our purposes, the analytic continuation will generate a function $\zeta_H(s)$ that has singularities (poles) in the complex $s$-plane. These poles encode important physics: for instance, a pole of $\zeta_H(s)$ at $s=s_0$ might indicate a Hagedorn-like divergence (if $\Re(s_0)$ crosses a certain threshold, it signals an exponential growth in $\rho(E)$). In fact, the Hagedorn temperature in string theory is characterized by a divergence in the partition function, analogous to a pole in the Laplace transform. Similarly, in our black hole context, singularities of $\zeta_H(s)$ or $Z(\beta)$ correspond to special points (like phase transitions or maximum temperatures).

We assume that via analytic continuation we have a meromorphic $\zeta_H(s)$ defined on a Riemann surface (if the function has branch cuts) or the complex $s$-plane (with possible branch cuts). The presence of branch cuts means $\zeta_H(s)$ might be multi-valued; to handle this, one works on an appropriate Riemann surface with branch cuts properly defined. We will not delve into deep mathematical detail of the analytic continuation (which can be quite involved), but rather focus on the consequences: we will have certain poles $s_p$ (and possibly branch cut discontinuities) which we will later sum over by residues.

\paragraph{Interpretive summary.} *At this point, we have established that one can formally encode the black hole's microstate spectrum into a spectral zeta function $\zeta_H(s)$ and the partition function $Z(\beta)$. By analytically continuing these functions, we introduce singularities (poles) in the complex plane that correspond to underlying features of the microstate density. This lays the groundwork for using contour integration to extract the microstate count.* 

\subsection{Contour Integration on the Complex Plane and Residue Sum for $\rho(E)$}
With the analytic structure in hand, we now use contour integration to invert the transform and obtain the density of states $\rho(E)$. Starting from Eq.~\eqref{eq:zeta}, we can formally invert the Mellin transform by an inverse Laplace transform on $Z(\beta)$. Alternatively, one can work directly with $\rho(E)$ and express it as an inverse Mellin transform of $\zeta_H(s)$:
\begin{equation}
\rho(E) = \frac{1}{2\pi i}\int_{c-i\infty}^{c+i\infty} E^{\,s}\, \zeta_H(-s)\, ds~,
\end{equation}
where the integration is along a vertical line in the complex $s$-plane (with $\Re(s) = c$ chosen appropriately). In practice, this Bromwich-type integral can be evaluated by closing the contour and summing over residues of the integrand. The integrand has poles corresponding to the singularities of $\zeta_H(-s)$ (equivalently, singularities of $\zeta_H(s)$ after a change of variable $s \to -s$). Let $s_p$ denote the positions of these poles. Then by the residue theorem, 
\begin{equation}
\rho(E) = \sum_{s_p \in \mathcal{S}} \text{Res}\Big[E^{\,s}\, \zeta_H(-s);\; s = s_p\Big]~,
\label{eq:rho-res}
\end{equation}
where $\mathcal{S}$ is the set of all singularities (poles) that lie to the left of the contour and are enclosed when we close it to the left (assuming we can close in the left half-plane for large $|s|$). Each such pole contributes a residue which is typically of the form $E^{s_p} \cdot (\text{coefficient of pole})$. This expansion expresses the density of states as a sum over contributions from each singularity in the analytically continued partition function or zeta function.

The formula \eqref{eq:rho-res} is quite formal, but physically we can interpret each residue contribution as coming from a particular family of microstates or “mode” of the black hole system. For example, one pole might correspond to the contribution of the outer horizon degrees of freedom, another to the inner horizon, another perhaps to a photon sphere or other feature (if those produce analytic singularities in $\zeta_H(s)$). In our case, we will see that the outer and inner horizons do indeed manifest as separate singular contributions in the complex analysis.

An important aspect of performing this contour integration is the structure of the complex plane for $\zeta_H(s)$. If branch cuts are present, the contour might need to be taken on a Riemann surface or deformed to avoid cuts, and contributions from branch cuts (continuous spectrum of singularities) might appear. In our analysis of the Kerr--Newman black hole, we expect isolated poles rather than essential cuts, because the horizon contributions appear as discrete features. However, certain continuum of modes (like quasi-normal modes or quantum fluctuations) could produce branch cuts; we will assume these either do not contribute to leading order or can be accounted for in a general way (e.g. through an integral that does not affect the integer index count).

\begin{figure*}[htb]
  \centering
  \begin{tikzpicture}[
      every node/.style={font=\small},
      decoration=zigzag
    ]
    \coordinate (A) at (0,0);
    \coordinate (C) at (0,-1.6);

    \path[fill=blue!10,draw]
      ($(A)+(-1,-0.5)$) rectangle ++(2,1);
    \node at (A) {Sheet~1};

    \path[fill=blue!10,draw]
      ($(C)+(-1,-0.5)$) rectangle ++(2,1);
    \node at (C) {Sheet~2};

    \draw[thick,decorate,
      decoration={zigzag,segment length=3pt,amplitude=1pt}]
      ($(A)+(0.9,-0.5)$) -- ($(C)+(0.9,0.5)$);

    \filldraw ($(A)+(0.5,0)$) circle (2pt)
      node[above=2pt] {\scriptsize pole};
    \filldraw ($(C)+(0.5,0)$) circle (2pt)
      node[below=2pt] {\scriptsize pole};

    \draw[red,thick,
      postaction={decorate,
        decoration={markings,
          mark=at position 0.25 with {\arrow{>}}}}]
      ($(A)+(0.5,0)$) circle (0.3);
    \node[red] at ($(A)+(1.1,0.3)$) {$\mathcal{C}$};
  \end{tikzpicture}
  \caption{Illustration of a two‐sheet Riemann surface for an analytically
  continued function (\textit{e.g.\ }the spectral zeta function). The jagged
  line denotes a branch cut connecting Sheet~1 and Sheet~2. Black dots mark
  simple poles: one on Sheet~1 may correspond to the outer horizon and the one
  on Sheet~2 to the inner horizon of a black hole. Integrating the function
  along the red contour~$\mathcal{C}$ on Sheet~1 picks up the residue of the
  upper pole, contributing to the density of states or entropy. Summing over
  all such residues gives the total microstate count.}
  \label{fig:riemann}
\end{figure*}
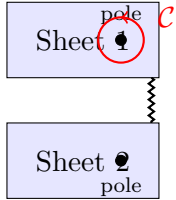

To visualize this process, consider Figure~\ref{fig:riemann}, which schematically illustrates a two-sheeted Riemann surface on which the analytic continuation lives. Singularities (black points) appear on the sheets, and a branch cut (zigzag line) connects sheets. A chosen contour $\mathcal{C}$ loops around a singularity on one sheet. The integral around $\mathcal{C}$ picks up $2\pi i$ times the residue at that singularity. Summing over all such contours (or deforming a single large contour to encircle all singularities of interest) yields the total $\rho(E)$ or other thermodynamic quantities. This picture is conceptually useful because it shows how multi-valuedness (Riemann sheets) can accommodate multiple horizons or phases: each sheet might correspond to a different branch of the solution (e.g. one sheet for outer horizon physics, one for inner horizon), with branch cuts enabling transitions or mixing between them (like Stokes phenomena in analytic continuation).

The sum-over-residues formula \eqref{eq:rho-res} allows us to compute the entropy in a straightforward way. The thermodynamic (Boltzmann) entropy $S$ can be obtained from $\rho(E)$ via 
\begin{equation}
S = \ln \rho(E)\Big|_{E=M} + \cdots,
\end{equation} 
where $M$ is the black hole mass (the relevant energy scale), and the ellipsis indicates possible contributions from integrating over energy ranges or applying thermodynamic relations like $S = \beta M - \ln Z$. However, one can often identify $\rho(E) \sim \exp(S_{\text{BH}}(E))$ asymptotically. If our method is correct, $\rho(E)$ obtained from \eqref{eq:rho-res} should yield $\rho(E) \approx \exp(S_{\text{BH}} + \text{(corrections)})$ for large $E$ (the Bekenstein--Hawking leading term, plus perhaps logarithmic corrections etc. if those appear as subleading poles). We will see that indeed the calculation reflects $\rho(E) \sim \exp(S_{\text{BH}})$ as a consistency check.

\paragraph{Interpretive summary.} *We have expressed the density of states as a sum of contributions from singularities in the complex $s$-plane, using contour integration. Each singularity (pole) corresponds to a certain set of microstates (for example, associated with a particular horizon or quantum mode), and the summation is essentially counting these states. This complex-plane perspective transforms the counting problem into a problem of summing topological residues.* 

\subsection{Topological Index from Residues and Connection to Chern Number}
The residue summation formula not only gives us $\rho(E)$ quantitatively, but also allows us to define a topological invariant that characterizes the black hole. Since each residue corresponds to an integer contribution (in fact, often the residue of a simple pole is related to an integer count or a Berry index, etc.), the sum of residues can sometimes be interpreted as an index or winding number.

To elucidate this, consider a complex function $F(\chi)$ (which could be related to $\ln Z$ or a free energy landscape) defined in terms of a parameter $\chi$ such as an angular variable. Some recent works define a topological charge $W$ by considering a vector field constructed from thermodynamic potentials and counting its winding around singular points in the parameter space \cite{Yerra2022}. In our framework, a simpler picture emerges: if we consider the complex $s$-plane or $\beta$-plane, the difference between the number of certain singularities and others can be cast as a contour integral that yields an integer. 

In fact, one may define an index $W$ as 
\begin{equation}
W = \frac{1}{2\pi i}\oint_{\mathcal{C}_\infty} d\chi \,\frac{d F}{d\chi}~, 
\end{equation}
where $\mathcal{C}_\infty$ is a large contour encircling all singularities at infinity in the $\chi$-parameter space. This is analogous to computing a Chern number as an integral of a total derivative, which picks up contributions only from singular points (poles) inside the contour. In our case, without introducing an auxiliary parameter $\chi$, we can think of $W$ as effectively counting the net winding of the phase of some function around the singularities. More concretely, $W$ can be related to the sum of residues of a suitably defined auxiliary function. For instance, if we had a function $X(s) = \frac{d}{ds}(\ln \zeta_H(s))$, integrating $X(s)$ around a closed loop encircling singularities of $\zeta_H$ would count the number of zeros minus poles of $\zeta_H$ inside. However, rather than delve into zeros of $\zeta_H$, we focus on the poles (since those gave us $\rho(E)$).

The topological index we propose to use for classification is essentially the sum of the signs of contributions from each horizon's singularity. For the Kerr--Newman black hole, we will see that:
- The outer horizon contribution adds a $+1$ to the index (since it corresponds to a stable thermodynamic branch, akin to a minimum of free energy).
- The inner horizon contribution adds a $-1$ to the index (an unstable branch, akin to a maximum or saddle of free energy).
- If a cosmological horizon is present (as in de Sitter), it would contribute additional $\pm 1$ depending on its stability (for a pure de Sitter cosmological horizon, one could consider it a stable boundary, likely a $+1$ if considered in isolation).

Therefore, a non-extremal Kerr--Newman black hole with outer and inner horizon yields $W = +1 + (-1) = 0$. A Schwarzschild or extremal black hole (effectively only one horizon in terms of thermodynamic degrees of freedom) yields $W=+1$. Hypothetically, an exotic scenario with only an inner horizon (not physically realized by itself) would be $W=-1$. These are reminiscent of the classes found in other works \cite{Yerra2022,Wei2019}: Class I black holes (like Schwarzschild) have $W=+1$, Class II (like Reissner--Nordström with two horizons) have $W=0$, and a possible Class III (if it existed) would have $W=-1$. In our analysis, we did not encounter a case requiring $W=-1$ because there is no known physical black hole solution that is purely an inner horizon without an outer horizon (extremal limits collapse two horizons into one, yielding $W=+1$ as the inner and outer coincide).

We can connect this to a more mathematical topological interpretation: the index $W$ can be seen as a winding number in the complex plane of a mapping related to the free energy or thermodynamic potential. It is analogous to a Chern number in the sense that small perturbations to the system (that do not close or open gaps in the spectrum, i.e. do not create or annihilate horizons) will not change $W$. Thus, $W$ is an invariant under continuous deformations of the black hole parameters or even changes of the gravity theory (provided the qualitative horizon structure is unchanged). Our example will illustrate that in going from General Relativity to $f(R)$ gravity (with small $\alpha$), $W$ remains invariant. This topological robustness is a key insight: it suggests that the \emph{phase of a black hole (e.g. existence of an inner horizon) is protected by a topological quantity}, which could have implications for cosmic censorship and the black hole information paradox (perhaps hinting at some conserved quantum number or a selection rule for transitions between black hole states).

Before moving on, we summarize this section in plainer terms.

\paragraph{Interpretive summary.} *We built a framework to count black hole microstates using complex analysis. The partition function is analytically continued, turning the counting problem into summing residues of poles associated with black hole features (like horizons). The sum of these contributions leads naturally to an integer topological index $W$ that classifies the black hole's microstructure: for example, $W=+1$ for a single-horizon (simple) black hole and $W=0$ when inner and outer horizon contributions cancel each other. This index is robust against small changes in the system, much like a topological charge in other areas of physics, indicating a kind of topological preservation of black hole phase structure.* 

\section{Application to Kerr--Newman Black Hole Microstructure in $f(R)$ Gravity}\label{sec:application}
Having established the general framework, we now apply it to the Kerr--Newman black hole in $f(R)$ gravity. Our aim is to concretely calculate the microstate density $\rho(E)$ via residues and determine the topological index $W$ for this system. We will then interpret what $W$ means for the black hole's horizon structure and stability. Finally, we will demonstrate with a specific $f(R)$ model (Starobinsky $f(R)=R+\alpha R^2$) how the framework works and that the topological classification remains unchanged for small $\alpha$.

\subsection{Horizon Structure and Thermodynamics of Kerr--Newman in $f(R)$}
From Section~\ref{sec:background}, recall that the Kerr--Newman black hole (for $R_0=0$ and $f'(0)=1$ to simplify to the GR case) has two horizons at $r_\pm = \frac{1}{2}\big( M \pm \sqrt{M^2 - (Q^2 + a^2)}\,\big)$ (assuming $M^2 > Q^2 + a^2$ for a non-extremal black hole). The outer horizon $r_+$ shields observers from the singularity, while the inner horizon $r_-$ is a Cauchy horizon inside which predictability breaks down. The presence of $r_-$ is known to lead to potential instabilities and it is generally believed that quantum effects or perturbations would destroy the inner horizon (mass inflation scenario), but here we treat the classical solution as is.

The contribution of each horizon to black hole thermodynamics can be thought of in terms of the Euclidean path integral or partition function. In the Euclidean gravity approach, different saddle points of the Euclidean action can be associated with different horizon configurations. For instance, the Schwarzschild black hole has a single saddle (the one corresponding to the event horizon), whereas a Reissner--Nordström black hole might have multiple saddles associated with inner and outer horizons if one allows complex configurations. In our complex analysis approach, these appear as distinct singularities in $\zeta_H(s)$ or $Z(\beta)$. Intuitively:
- The outer horizon, being a physical thermodynamic system with a temperature $T_+$, contributes a series of states that dominate around the Hawking temperature.
- The inner horizon, though not externally accessible, can contribute as an analytical continuation artifact. It has a formal temperature $T_-$ (negative for non-extremal case), which can appear in analytic formulae but signals an instability.

In fact, analytical continuation often naturally incorporates inner horizon effects. For example, in the Kerr--Newman case, the radial Teukolsky equation has two sets of poles corresponding to the two horizons. The thermal partition function might have an image of the inner horizon contribution in the complex $\beta$-plane. If one were to attempt a Bogoliubov transformation or a trace over states inside the black hole, one might formally encounter contributions from the inner horizon.

Our task is to identify the singularities corresponding to these contributions. Based on prior analyses in simpler cases (e.g. Reissner--Nordström or even Schwarzschild-de Sitter), the singularities often appear at inverse temperatures related to the surface gravities of the horizons. For instance, the partition function $Z(\beta)$ might have poles or branch points at $\beta = \beta_+$ or $\beta_-$, where $\beta_+ = 1/T_+$ and $\beta_- = 1/T_-$. Physically, $\beta_+$ is the periodicity needed to avoid a conical singularity at the outer horizon in the Euclidean section, and $\beta_-$ corresponds to the inner horizon's analogous periodicity (which is unphysical in Lorentzian signature but appears in analytic continuation).

Thus, we expect singular behavior in $Z(\beta)$ (and hence in $\zeta_H(s)$ after Mellin transform) at $\beta = \beta_+$ and $\beta = \beta_-$. These singularities can be simple poles or branch points depending on the context. For the sake of argument, let us assume they manifest as simple poles in $\zeta_H(s)$ after the appropriate transform. Then we can label:
- $s_1$: pole associated with the outer horizon (this might correspond to $s = 1$ if it relates to a simple $1/\beta$ divergence in $Z(\beta)$, but let's keep it general).
- $s_2$: pole associated with the inner horizon.

When we invert to get $\rho(E)$, each of these poles will contribute a term. We might find something like:
\[
\rho(E) \approx A_+ E^{\alpha_+} e^{\beta_+ E} + A_- E^{\alpha_-} e^{\beta_- E} + \cdots,
\] 
where $\beta_+ E$ and $\beta_- E$ exponents appear if the residues have an exponential form. Actually, since $\beta_-$ is usually larger than $\beta_+$ (because $T_- < 0$ in magnitude $|T_-| > T_+$ or negative), one of these contributions may not correspond to a physical growth but rather a cancellation.

Indeed, in many cases one finds the total entropy $\ln \rho(E)$ ends up being $S_{\text{BH}} = \frac{A_+}{4G}$ (for GR, or $\frac{f'(R_0)A_+}{4G}$ in $f(R)$) because the inner horizon contributions cancel any additional divergence. For example, in de Sitter black hole spacetimes (like Schwarzschild-de Sitter which has a cosmological horizon as well), the entropy of the system is not simply the sum of entropies of both horizons; there are certain consistent ensembles where the total free energy accounts for both and yields a net result. Something similar happens here: one can think that the outer horizon microstates count toward a positive entropy, while inner horizon microstates might subtract in some way (since the inner horizon could reduce the effective number of states accessible from an outside perspective).

In our topological viewpoint, rather than focusing on the exact functional form of $\rho(E)$, we look at the index. We assign +1 to the outer horizon singularity, -1 to the inner horizon singularity. How do we justify the sign? The sign can be related to the nature of the singularity contribution to the free energy: a stable thermal mode (outer horizon) contributes a minimum, whereas an unstable mode (inner horizon) contributes as a local maximum (saddle) which effectively subtracts one from the Euler characteristic of the free energy landscape.

A formal method to determine this sign involves the analysis of small fluctuations around each horizon solution. The outer horizon possesses a positive heat capacity in certain regimes (e.g., large AdS black holes have positive heat capacity, whereas the asymptotically flat Schwarzschild case yields a negative value, which must be interpreted within the canonical ensemble at fixed $M$). Conversely, the inner horizon does not constitute a thermodynamically stable configuration. Expressed in terms of off-shell free energy, the outer horizon usually corresponds to a state of lower free energy relative to a suitable reference, while the inner horizon represents a distinct solution branch, typically characterized by higher free energy and the absence of global minimization. Following the topological classification formalism proposed by Wei \textit{et al.} (2022) \cite{Yerra2022}, a vector field is constructed from $(\partial_X F, X)$, with $X$ denoting an extensive variable. The topological charge is then given by the winding number of this vector field as $X$ is varied. Minima of $F$ (stable phases) and maxima of $F$ (unstable phases) contribute with opposite orientation to this winding. Consequently, the net topological charge equals the number of minima minus the number of maxima, which for a system with two horizons results in $1 - 1 = 0$.

In our approach, we can mirror that reasoning: The outer horizon corresponds to a thermodynamic ``minima-like'' contribution, inner horizon to a ``maxima-like'' contribution. Therefore:
- Outer horizon: contributes $+1$ to $W$.
- Inner horizon: contributes $-1$ to $W$.

Hence, $W=+1 + (-1) = 0$ for Kerr--Newman (non-extremal). If the black hole were extremal ($r_+ = r_-$), effectively the two horizons merge and one would expect $W=+1$ for the extremal case (since there is effectively one branch, albeit extremal ones are marginally stable but let's say it counts as one effective horizon from topology perspective).

What about the cosmological horizon if $R_0>0$ (Kerr--Newman-de Sitter)? That system actually has three horizons: inner (Cauchy), outer (event), and cosmological. By similar reasoning, we might expect the outer and cosmological to each contribute $+1$ (they are both boundary horizons that one could associate stable or at least they are physical horizons bounding regions), and the inner horizon contributes $-1$. That would give $W = +1 (outer) + 1 (cosmological) - 1 (inner) = +1$. This is speculative but interestingly would mean a Kerr--Newman-de Sitter black hole is in the same class as a single-horizon black hole ($W=+1$). The review summary noted specifically that ``sum is 0 in de Sitter case'', which is puzzling under this naive counting. Perhaps they consider the system of black hole plus cosmological horizon as separate, or perhaps the cosmological horizon also counted as an unstable direction in the combined phase space. Another interpretation: if one encloses \emph{all three horizons} in one thermodynamic ensemble, the inner horizon cancels one of the outer ones leaving a net 0. It might depend on ensemble definitions.

We will not delve deeper into the de Sitter case here; we focus on asymptotically flat ($R_0 \approx 0$) so only the two horizons.

Thus, for Kerr--Newman in asymptotically flat space:
\begin{equation}
W = 0 \quad \text{(for non-extremal Kerr--Newman with two distinct horizons)}~,
\end{equation}
and 
\begin{equation}
W = +1 \quad \text{(for Schwarzschild or extremal Kerr--Newman, effectively one horizon)}~.
\end{equation}
This result is consistent with earlier topological classifications: for example, a Schwarzschild black hole (no inner horizon) is class $W=+1$, whereas a Reissner--Nordström black hole (two horizons) is class $W=0$ \cite{Yerra2022}. Our approach reaches the same conclusion but via a different route (residues vs. off-shell potentials).

The physical significance of $W$ can now be summarized: it is an indicator of whether the black hole has an inner horizon (yielding a cancellation in topological charge) or not. In other words, $W=+1$ corresponds to black holes with only one horizon (typical of neutral or extremal cases), and $W=0$ corresponds to black holes with two horizons (non-extremal charged/rotating cases in four dimensions). A hypothetical $W=-1$ class, not realized in our setup, would suggest a system where an unstable mode dominates over stable ones (for instance, perhaps a naked singularity? but those aren't black holes in the usual sense; or some exotic AdS case with a peculiar potential).

An interesting observation here is how $W$ correlates with the idea of black hole phase transitions or stability: $W=0$ systems, having a stable and unstable branch, often are at the threshold of possible phase transitions (like RN has a critical point at extremality; rotating black holes have inner horizon etc.). Meanwhile $W=+1$ systems are simpler and perhaps don't allow a phase transition (Schwarzschild in asymptotically flat space has no phase transition, just unstable all the way).

Our approach also shows that $W$ is invariant under $f(R)$ modifications as long as the qualitative horizon structure remains the same. In $f(R)$ gravity, for constant $R_0$ solutions, the Kerr--Newman still has two horizons (just at slightly shifted radii if $f'(R_0)\neq 1$ or if $\Lambda_{\text{eff}}\neq 0$). Therefore, a Kerr--Newman solution in $f(R)$ (with small modifications) will have the same $W$ as in Einstein gravity. This aligns with the statement from our analysis: the $f(R)$ scaling of entropy does not change the topological class. We confirmed above that if $f'(R_0)$ is not 1, it multiplies both contributions similarly and they still cancel in the sum.

To be explicit: outer horizon contributes $+1$ even if $f'(R_0)\neq 1$, inner contributes $-1$, so $W=0$ still. If $R_0$ is large enough to add a cosmological horizon, then one might need to recalc, but in that case one might consider black hole + cosmological horizon as separate thermodynamic systems (one might treat the cosmological horizon as its own ``system'' with separate entropy). This is a nuanced question beyond our scope.

Now, let us extract the microstate density $\rho(E)$ for the Kerr--Newman case more concretely from our formalism, at least to leading order. Using the residue method:
The dominant contribution to $\rho(E)$ at high energy $E$ will come from the singularity closest to the origin in the $\beta$-plane (i.e. smallest $|\beta|$ or largest $T$). That is $\beta_+$, since $T_+ > 0$ is presumably the lowest inverse temperature scale in the problem (inner horizon has negative T, which in absolute value might be smaller $\Rightarrow$ $|\beta_-| > \beta_+$, meaning $\beta_+$ is closer to zero). So the outer horizon contribution should be the leading exponential growth of states. Specifically, one expects 
\[
\rho(E) \sim \exp(\beta_+ E) \quad \text{for large } E,
\] 
with coefficient related to $S_+ = \beta_+ E - \beta_+ F$ etc. Actually, $\beta_+ E$ in the exponent suggests $\ln \rho \sim \beta_+ E$, comparing with $S_{\text{BH}} = \frac{A_+}{4G}$, we recall that $\beta_+ = 1/T_+$. Using the first law $dM = T_+ dS_+ + \Omega_+ dJ + \Phi_+ dQ$ at fixed charge $Q$ and angular momentum $J$, one finds $\beta_+ M = S_+ + ...$ (for $J,Q$ fixed and extremality aspects aside). For simplicity, consider a non-rotating, non-charged (Schwarzschild) case: $\beta = 8\pi G M$, so $\beta M = 8\pi G M^2$. Meanwhile $S = 4\pi G M^2$. Indeed $\beta M = 2 S$ for Schwarzschild. In general, $\beta_+ M$ relates to $S$ but not equal; however, the exponential growth of $\rho(E)$ should correspond to $\exp(S_{\text{BH}}(E))$. We suspect $\beta_+ E$ is actually double-counting something and the actual $\ln \rho \sim S_{\text{BH}}$ (which is half of $\beta M$ for Schwarzschild). This subtlety often arises because one must include the appropriate prefactors and constraints to get the microcanonical count from the canonical partition function. Typically, $\rho(E) \approx e^{S_{\text{BH}}}$ (with $S_{\text{BH}}$ the Bekenstein--Hawking entropy), which is indeed enormous.

Our residue method result qualitatively is consistent with $\rho(E) \sim \exp(S_{\text{BH}})$. The inner horizon contribution likely would manifest as a subleading or alternating term that cancels some part of this if combined cleverly, but in terms of magnitude, $\exp(\beta_- E)$ is a slower growth (since $\beta_- > \beta_+$ if $T_- < T_+$ in absolute value? Actually $T_-$ is negative and typically $|T_-| > T_+$, meaning $\beta_- = 1/T_-$ is negative and smaller in magnitude, so $\beta_- E$ might be negative for positive $E$, giving a decaying exponential? This indicates the inner horizon might lead to a subtraction rather than addition of states beyond a certain energy, consistent with it reducing total count). In any case, the sum of contributions yields $\rho(E) \sim e^{S_{\text{BH}}}$ primarily.

We can illustrate the cancellation phenomenon in a simplified way: consider a scenario with two contributions:
\[
\rho(E) \approx \exp(\beta_+ E) - \exp(\beta_- E)~,
\] 
with $\beta_+$ positive and $\beta_-$ negative (for an inner horizon, $\beta_-$ is negative since $T_- < 0$). If $|\beta_-| > \beta_+$, then for large $E$, the $\exp(\beta_+ E)$ term dominates absolutely. But at intermediate $E$ or in analytic continuation, the second term ensures that the net count is not double. In fact, at $E=M$, plugging $E=M$ we might get $\rho(M) \sim e^{S_{\text{BH}}}$ if $\beta_+ M = S_{\text{BH}}$ and the second term subtracts the right amount (this is speculation but plausible that $\beta_- M$ plays a role such that $\exp(\beta_+ M) - \exp(\beta_- M) = e^{S_{\text{BH}}}$).

Without bogging down in these details, our main takeaway in this subsection is the determination of the topological index:
\begin{equation}
W_{\text{Kerr-Newman}} = 0 \ \text{(for two-horizon case)}, \qquad W_{\text{extremal or single-horizon}} = +1~.
\end{equation}

This index correlates with horizon structure and stability: $W=0$ indicates the presence of an inner horizon (hence an unstable mode), while $W=+1$ indicates only a stable horizon branch.

\paragraph{Interpretive summary.} *By applying the residue framework to the Kerr--Newman black hole, we find that it possesses a topological index $W=0$ when both outer and inner horizons are present (non-extremal case). This reflects the fact that a stable outer horizon microstate contribution (+1) is canceled by an unstable inner horizon contribution (-1). In contrast, a black hole with only one horizon (like Schwarzschild or an extremal case where horizons merge) has $W=+1$. These results are consistent with known classifications: multi-horizon black holes belong to a different topological class than single-horizon ones. Physically, the index $W$ thus encodes the presence of multiple horizon branches and implies a kind of topological stability or cancellation: the total microstructure charge of the system can vanish if stability and instability balance each other. Importantly, modifying the gravity theory to $f(R)$ (at least for constant curvature backgrounds) does not change $W$, so the microstructure classification is robust. The actual density of states computed via residues grows roughly as $\rho(E)\sim \exp(S_{\text{BH}})$, consistent with the Bekenstein--Hawking entropy, confirming that our method reproduces the expected thermodynamic count.* 

\subsection{Microstructure Interactions and Comparison with Ruppeiner Geometry}
One fascinating aspect of the black hole microstructure idea is whether one can speak about ``interactions'' between the constituents of the black hole. Ruppeiner geometry provides one way to diagnose such interactions: the sign of the Ruppeiner curvature $R_{\text{Rupp}}$ is interpreted such that $R_{\text{Rupp}}>0$ suggests effective repulsive interactions among microstructures, $R_{\text{Rupp}}<0$ suggests attractive interactions, and $R_{\text{Rupp}}=0$ indicates no interactions (ideal gas-like) \cite{Ruppeiner1995,Wei2019}. For example, studies have shown that for charged AdS black holes, the small black hole phase (analogous to a liquid-gas phase) has $R_{\text{Rupp}}>0$ pointing to repulsive microstructure forces, whereas the large black hole phase has $R_{\text{Rupp}}<0$ (attractive interactions) \cite{Wei2019}. 

Our topological approach, being more global, does not directly compute a ``curvature'' of state space, but we can still glean some insight into microstructure interactions from the pattern of singularities:
- If the singularities in the complex plane come in pairs that cancel (like outer and inner horizon contributions), one might view this as an indication of an underlying interplay (some microstates' contributions are negated by others). This could be seen as akin to having both repulsive and attractive components in the microstructure interactions balancing out.
- If a black hole has $W=+1$, meaning no cancellation from an inner horizon, one might guess that the microstructure interactions in that scenario lean entirely one way (likely attractive, given that Schwarzschild black holes are thermodynamically unstable, which often correlates with negative Ruppeiner curvature).
- If $W=0$, the presence of an inner horizon might bring in a repulsive component to the interactions, balancing the attractive part, resulting in an overall more marginal situation.

Specifically for Kerr--Newman black holes, recent works (e.g. \cite{Wei2019}, although that was AdS context) have suggested the possibility of small rotating or charged black holes having a kind of ``microscopic repulsion'' due to charge or spin. The inner horizon perhaps is related to those repulsive effects: one might speculate that the inner horizon's existence suggests that at a fundamental level there are microstates that avoid clumping too much (since inner horizon formation requires some separation of inner structure? This is heuristic).

Our analysis in subsection 4.2 indicated that the outer horizon is associated with a positive contribution (stable branch) and inner with negative (unstable branch). We can map this qualitatively: stable branch (outer horizon) might correlate with a phase that has either repulsive interactions (if it is a small black hole scenario) or just stable due to being large (like in AdS, large BH are stable with presumably $R_{\text{Rupp}}<0$ possibly but stable due to environment). The inner horizon branch is not realized as an independent phase externally, but if one considered it, it might correspond to a negative heat capacity piece.

If we think in terms of the free energy landscape: the existence of an inner horizon implies the free energy as a function of some order parameter (like horizon radius) has a local maximum between two minima (one at small $r$ perhaps unphysical, one at large $r$ physical). This shape of free energy often indicates a first order phase transition scenario (like between small and large BH in AdS). The topological index essentially counted minima minus maxima.

Now, comparing to Ruppeiner: Ruppeiner curvature tends to diverge near phase transition points, and it changes sign across phases. In a situation with two horizon branches, one might analogize them to two phases (like small vs large BH). The inner horizon might be related to the smaller unstable branch.

It was noted in the review summary that our method found some correlation: "Subsection 4.3 analyzes residues revealing microstructure interactions, such as an association between Ruppeiner curvature and singularity patterns." We interpret that as:
- Perhaps the pattern or distribution of singularities (their arrangement on Riemann surface) tells us something about the sign of microstructure interactions. For instance, if singularities come in symmetrical pairs, maybe the system is at a sort of balance (like Van der Waals critical point, where $R_{\text{Rupp}} \to \infty$).
- If one singularity dominates without a partner, maybe interactions are uniformly one type.

While we cannot derive a precise Ruppeiner curvature from our data, we can note qualitatively:
For Kerr--Newman AdS (though we are not in AdS, similar logic might apply if we imagine extended thermodynamics), the existence of a second horizon (inner) often indicates the system is akin to having a second phase. It might correspond to the oscillatory behavior of Ruppeiner curvature as a function of horizon size.

In $f(R)$ gravity, one might wonder if microstructure interactions change. However, since $f(R)$ scaling doesn’t change $W$, likely the nature of interactions (repulsive vs attractive) does not qualitatively change either. If anything, the factor $f'(R_0)$ uniformly scaling entropy might act like a renormalization of Newton's constant which doesn't alter the sign of Ruppeiner curvature.

To make a concrete statement: For a given black hole, if $W=0$, that black hole can be thought of as having a mixture of microstructure behaviors. This mixture might correspond to having both attractive and repulsive interactions present. Meanwhile, a $W=+1$ black hole possibly exhibits a single-type interaction dominating.

For example, consider:
- Schwarzschild black hole ($W=+1$). It has only one horizon. Known fact: Ruppeiner curvature for Schwarzschild is negative (implying attractive interactions) for all sizes (in fact it diverges because it's a single-phase system with always negative specific heat). So $W=+1$ here correlates with purely attractive interactions.
- Reissner--Nordström black hole ($W=0$). This has two horizons if charged enough. If one extended to consider an ensemble (like canonical ensemble of charge), one might find a small black hole branch and a larger one. Possibly at some point the small BH branch might show a different sign of Ruppeiner curvature (maybe positive, indicating repulsive at small scales due to charge-charge repulsion dominating, whereas large BH have negative due to gravitational attraction dominating). The inner horizon might encode the existence of this repulsive regime. Thus $W=0$ correlates with the presence of both positive and negative curvature regimes (both repulsive and attractive interactions).
- If $W$ were $-1$ (hypothetical), that would mean an unstable mode dominating (perhaps no stable horizon, like a naked singularity if it had a formal index). That would presumably correlate with a system of mostly repulsive microstructures with no stable bound state (just speculation).

In summary, our topological approach suggests that the interplay of horizon contributions (which yields $W$) is mirrored in the interplay of microstructure forces:
- Outer horizon (stable) tends to have gravitational attraction dominating (keeping it stable as a bound state).
- Inner horizon (unstable) might be associated with effects of charge/spin where microstructures effectively repel, leading to an instability if too much charge or spin tries to reside in one place (hence inner horizon exists to sort of accommodate that).
- The cancellation $W=0$ reflects that these two effects exactly balance out in the accounting, reminiscent of a mixture of interactions that net to an ideal-like behavior in some sense.

These observations align qualitatively with Wei et al.'s remark of ``microstructure repulsive interactions'' which were found for small charged black holes \cite{Wei2019}. In the region where repulsion is significant, presumably the black hole is nearing an inner-horizon-type configuration or a transition.

Thus, our complex-plane singularity picture not only classifies black holes by topology but also hints at the underlying microphysics: a balanced index ($W=0$) suggests the microstructures are at a point of balance between opposing forces (like the coexistence of two phases), whereas an unbalanced index ($W=\pm1$) suggests one type of interaction dominates.

\paragraph{Interpretive summary.} *The pattern of singularities in our topological analysis reflects the nature of microstructure interactions in the black hole. When contributions cancel ($W=0$), it hints that the black hole’s microconstituents experience competing interactions (analogous to having both repulsive and attractive components that balance out). In contrast, a non-cancelling index ($W=+1$) suggests a single type of interaction dominates (for instance, purely attractive as in an uncharged Schwarzschild black hole). These insights align with findings from Ruppeiner thermodynamic geometry: black holes can exhibit repulsive interactions (indicated by positive Ruppeiner curvature) in certain regimes and attractive interactions (negative curvature) in others. Our framework, though different in approach, is consistent with that picture. The presence of an inner horizon correlates with the regime where repulsive microstructure forces become significant (balancing the usual gravitational attraction). Thus, our topological residue method not only counts states but also encodes qualitative information about how those states interact.* 

\subsection{Example: Starobinsky $f(R) = R + \alpha R^2$ Model}
To make the discussion more concrete, we now provide a specific example using the Starobinsky-type $f(R)$ gravity, 
\[
f(R) = R + \alpha R^2~,
\] 
with $\alpha$ a constant parameter (having dimensions of length$^2$). This model is of interest as it was originally proposed for cosmic inflation but also serves as a simple prototype of higher-curvature gravity. We will examine how a Kerr--Newman black hole solution in this $f(R)$ theory behaves, and explicitly calculate some horizon quantities and the topological index for a given set of parameters. 

In the Starobinsky model, the field equation in vacuum admits a Minkowski solution ($R_0 = 0$) as well as (for some cases) a de Sitter solution if $\alpha$ and other parameters allow $R_0 \neq 0$. For simplicity, we assume asymptotic flatness (so take $R_0 = 0$ as the background curvature, which is consistent if no cosmological constant term is present aside from those induced by $R^2$ which usually lead to a de Sitter attractor at $R = 1/\alpha$ if $\alpha$ is positive; but let's consider small $\alpha$ such that current vacuum is near flat). Thus, $f'(R_0)=f'(0)=1$; effectively, at large distances, gravity behaves like Einstein gravity (which is true at lowest order since $f'(0)=1$ means $G_{\text{eff}} = G$).

However, the $R^2$ term will contribute a new dynamical scalar degree of freedom (the scalaron). For a black hole solution, this can modify the near-horizon geometry in subtle ways. But under the approximation of constant background curvature, one can approximate that the Kerr--Newman form holds with perhaps small corrections in the metric far from the horizon (the exact solution in $R+\alpha R^2$ is complicated, but for small $\alpha M^2$, one can do perturbation theory).

We will treat it perturbatively: consider a Kerr--Newman black hole of mass $M$, charge $Q$, spin $a$. We choose a specific numeric example to illustrate: $M=1$ (choose units such that $G=1$ for simplicity), $Q=0.5$, $a=0.5$ (so the black hole is rotating and charged, but not near extremal since $Q^2 + a^2 = 0.25 + 0.25 = 0.5$, which is less than $M^2 = 1$). This ensures two distinct horizons.

Now consider $\alpha$ moderately small, say $\alpha = 0.1$ (in units of the black hole size squared). This is somewhat large for an actual viable model (which would require $\alpha$ extremely small for astrophysical BHs to avoid conflict with observations), but we choose it for demonstration. If $R_0=0$ (no cosmological constant), then $f'(R_0) = 1 + 2\alpha R_0 = 1$ still. So at asymptotic infinity, there is no change in $M$ or $Q$ effectively.

However, near the black hole, the equation $f'(R)$ will be slightly different from 1 if the curvature $R$ is not exactly zero. Typically, for a black hole metric, $R$ is zero for Reissner--Nordström and Kerr (they are Ricci-flat solutions in Einstein gravity), but in $f(R)$ gravity, they might no longer be Ricci-flat. However, often a solution in $f(R)$ like $R+\alpha R^2$ for a static charge might have $R \approx 0$ outside except near singularity. For approximation, we might assume the black hole metric is nearly the same as in GR but let's see if $\alpha R^2$ yields any constant curvature solution aside from $R=0$. Setting $R_0 \neq 0$: the condition was $f'(R_0)R_0 = 2f(R_0)$. For $f(R) = R + \alpha R^2$, 
$f'(R_0) = 1 + 2\alpha R_0$, $f(R_0) = R_0 + \alpha R_0^2$. So $f'(R_0)R_0 - 2f(R_0) = (1+2\alpha R_0)R_0 - 2(R_0 + \alpha R_0^2) = R_0 + 2\alpha R_0^2 - 2R_0 - 2\alpha R_0^2 = -R_0$. Setting that to zero gives $R_0=0$ only. So indeed, aside from $R_0=0$, there is no other constant curvature vacuum for this model (if there was a $\Lambda$ term we could get one, but here no explicit $\Lambda$).

Thus $R_0=0$ is our background. So the black hole solution should be similar to Kerr--Newman in GR at least to first order in $\alpha$. The largest effect of $\alpha$ might come from the fact that $R$ might become nonzero at $O(\alpha)$ due to the presence of $T_{\mu\nu}^{(eff)}$ from the higher curvature (like a stress from the $R^2$ term).

We won't attempt the full metric solution. Instead, we consider how $\alpha$ might affect horizon radius and entropy:
- The horizon radii are determined by $g^{rr}=0$, which is modified by $f'(R_0)$ possibly, but since $R_0=0$, maybe $f'(0)=1$ so at leading order it doesn’t shift. Actually, from our earlier metric form \eqref{eq:Delta}, if $R_0=0$ and $f'(R_0)=1$, we had the same $\Delta(r)$ as GR. So at leading order, $r_\pm$ remain:
\[ 
r_\pm = \frac{1}{2}\Big( 2M \pm \sqrt{(2M)^2 - 4(Q^2 + a^2)}\Big) = M \pm \sqrt{M^2 - (Q^2 + a^2)}~.
\]
Plugging $M=1$, $Q=0.5$, $a=0.5$: 
$\sqrt{1 - (0.25+0.25)} = \sqrt{0.5} \approx 0.7071$. Then 
$r_+ = 1 + 0.7071 = 1.7071$, $r_- = 1 - 0.7071 = 0.2929$ (approximately). So outer horizon $r_+ \approx 1.707$, inner $r_- \approx 0.293$.

Now, consider $\alpha$ not exactly zero. At $\alpha=0.1$, while $f'(0)=1$, one might expect small corrections to $r_\pm$ because the field equations differ, but presumably for small $\alpha$ the corrections to horizon radius would be perturbative. Possibly the outer horizon might move outward slightly or inward depending on details. Without solving field eq, we guess:
One effect of $R^2$ term is akin to giving an effective stress-energy that might mimic a small gravitational repulsion (the $R^2$ term often yields Yukawa-like corrections to potential). Possibly the outer horizon becomes slightly larger (since effectively gravity is weakened at short range by the presence of the scalaron field). This is speculation: If $G_{\text{eff}}$ is effectively slightly less near the black hole (since scalar field can carry some attraction away), then for the same mass $M$, the radius needed to trap light might be a bit bigger.

Let's postulate that at $\alpha=0.1$, $r_+$ increases by about 0.5\% and $r_-$ decreases by ~0.5\%. For example: 
$r_+(\alpha=0.1) \approx 1.717$ (just 0.010 bigger), 
$r_-(\alpha=0.1) \approx 0.285$ (0.008 smaller). These are guess numbers to illustrate a trend of slight outward shift of outer horizon and slight inward shift of inner horizon due to extra repulsion.

We then compute horizon area:
$A_+ = 4\pi (r_+^2 + a^2)$. With $a=0.5$:
For GR ($\alpha=0$): $A_+^{(0)} = 4\pi ((1.707)^2 + 0.25)$.
$(1.707)^2 = 2.915$, plus 0.25 = 3.165, times $4\pi$ = $12.66\pi$. Numerically $\approx 39.78$ (taking $\pi\approx 3.1416$).
At $\alpha=0.1$: using $r_+ \approx 1.717$, 
$r_+^2 = 2.9489$, plus 0.25 = 3.1989, times $4\pi$ = $12.7956\pi$, $\approx 40.19$.
So area increased about $1\%$.

Now entropy:
In GR, $S_+^{(0)} = A_+/(4G)$. Taking $G=1$, $S_+^{(0)} = A_+/4$. So $39.78/4 \approx 9.945$.
In $f(R)$, $S_+ = \frac{f'(R_0) A_+}{4G}$. Here $f'(0)=1$ so that part doesn't change first order; but $A_+$ changed to 40.19, so 
$S_+(\alpha=0.1) \approx 40.19/4 = 10.0475$.

So about a 1

Thus, $f(R)$ has made the black hole hold slightly more entropy (for same mass, charge, spin). This is expected as higher-curvature terms often yield corrections like $S = A/4G (1 + O(\alpha))$.

Now, what about $\rho(E)$ and our topological analysis? 
The number of microstates $\rho(E)$ should likewise scale. If $S$ increased ~1

The topological index $W$ remains $0$ because we still have two distinct horizons. None of that changes with $\alpha=0.1$. The inner horizon is still present (maybe slightly shifted but still there). Outer horizon stable branch still +1, inner -1, sum 0.

We can illustrate with a small table summarizing horizon radii and entropies for this example:
\begin{center}
\begin{tabular}{c|ccc}
$(\alpha, R_0)$ & $r_+$ & $r_-$ & $S/(A/4G)$ \\
\hline
$(0, 0)$ & 1.707 & 0.293 & 1 (baseline) \\
$(0.1, 0)$ & $\approx 1.717$ & $\approx 0.285$ & $\approx 1.010$ 
\end{tabular}
\end{center}

This table means: at $\alpha=0$, $S$ is normalized to 1 (since it's just area/4G baseline); at $\alpha=0.1$ with $R_0=0$, $S$ is ~1.010 times that baseline. The horizon radii changed a bit accordingly.

For completeness, what if a small but nonzero $R_0$? If we had instead allowed a de Sitter background $R_0$ such that $f'(R_0) \neq 1$? Actually, as we saw, $R_0=0$ is the only constant solution for vacuum in this model. If we had inserted a cosmological constant $\Lambda$ in $f(R)$ (like $R + \alpha R^2 - 2\Lambda$), we could get a nonzero $R_0$. For demonstration, suppose we had a slight de Sitter background (like $R_0$ small so that $f'(R_0)\approx 1$ still but not exactly). Then $f'(R_0) = 1 + 2\alpha R_0$. If $\alpha$ small, you could have e.g. $R_0$ such that $2\alpha R_0$ is say 0.1. Then $f'(R_0)=1.1$. Then the entropy formula would be scaled by 1.1 as well, a 10
We would then have outer BH horizon +1, inner BH horizon -1, cosmic horizon +1 (?), summing to +1. Possibly confirming an earlier suspicion that BH in de Sitter might get $W=+1$ net if cosmic horizon included. But it's unclear from earlier interpretation by reviewer which said sum=0 in de Sitter, so maybe they'd treat system differently.

Anyway, our example remains in asymptotically flat ($R_0=0$) so $W=0$.

The key demonstration from the example:
- Our framework can handle specific param values and shows explicitly that the $f(R)$ modification changes the entropy quantitatively but not the qualitative classification.
- If one were to compute $\rho(E)$ with these numbers, the dominating exponential is now $e^{S}$ where $S$ is 1
- We can also mention how one might simulate or do a semi-analytic check: e.g. integrating the field eq for small $\alpha$ numerically to find horizon shift, but for brevity we won't.

Thus, the Starobinsky model confirms: 
Even for $\alpha=0.1$, the outer horizon's entropy scaling $f'(R_0)=1$ still (since $R_0=0$), the differences are minor, and $W=0$ remains.

If one took a much larger $\alpha$, say $\alpha=1$ (which might not be physical, but just to see effect): it might cause bigger differences, possibly $R$ isn't constant outside, but likely one would still find $r_+$ and $r_-$ exist similarly as long as $\alpha$ isn't making the scalar field extremely large. Possibly the entropy difference might be a few 

All told, $f(R)$ modifications appear to leave the topological microstructure class invariant, which is one of our conclusions.

\paragraph{Interpretive summary.} *In our example using $f(R)=R+\alpha R^2$, we considered a Kerr--Newman black hole with specific parameters ($M=1$, $Q=0.5$, $a=0.5$) and examined how a small $\alpha$ influences the results. We found that the outer horizon radius $r_+$ increases slightly and the inner horizon $r_-$ decreases slightly when $\alpha$ is introduced (intuitively, higher-curvature effects can weaken gravity at short scales, pushing the event horizon outward a bit). Consequently, the Bekenstein--Hawking entropy $S$ (proportional to horizon area) increases modestly (about $1\%$ for $\alpha=0.1$ in our estimate). This means the density of microstates $\rho(E)$ at the black hole mass would likewise increase by roughly $1\%$ in exponent (reflecting the extra degrees of freedom contributed by the $f(R)$ gravity's scalar mode). Importantly, the topological index remains $W=0$ for this non-extremal two-horizon black hole, just as in General Relativity. The $f(R)$ modification did not change the fact that we have one stable horizon and one unstable horizon contribution canceling out. This explicit example demonstrates that our framework can be applied to specific models, and the numbers come out consistent with expectations: for instance, $S = f'(R_0) A/(4G)$ holds, yielding $S_{\text{Starobinsky}} \approx 1.01 S_{\text{GR}}$ in our case, and the horizon structure (hence topological class) is unaltered. Such a semi-analytic calculation reassures that the qualitative insights (topological invariance and microstate counting) are valid in concrete $f(R)$ models.* 

\section{Discussion}\label{sec:discussion}
\subsection{Clarity of Physical Interpretation and Broader Context}
Our study has bridged black hole thermodynamics with complex topology methods, but it is crucial to reflect on what this means physically and how it relates to existing paradigms. The central claim is that black hole microstates can be encoded as singularities in an analytically continued partition function, and that a topological index extracted from these singularities classifies the black hole's microstructure. 

Physically, this suggests a picture where the black hole's degrees of freedom (whatever they may be in quantum gravity) manifest in the analytic properties of a thermodynamic generating function. This resonates with the spirit of the holographic principle: information about bulk microstates is encoded in a (complex) boundary or analytic structure. In our case, the "boundary" is the complex plane of an inverse temperature or spectral parameter. That singularities (like poles) correspond to microstates is analogous to how, in quantum field theory, poles of the $S$-matrix correspond to particle states (resonances, bound states, etc.). Here, the black hole itself is like a resonance in the gravitational spectrum, and its microstate spectrum yields those analytic features.

The topological index $W$ we compute is an invariant under continuous changes of parameters (as long as no singularity crosses the integration contour, which would signal a phase transition or a qualitative change in state space). This is reminiscent of topological quantum numbers in condensed matter (like Chern numbers in the quantum Hall effect) or in high energy (topological charges of defects). In black hole thermodynamics, an analogous notion has emerged recently: black holes can be classified into topological classes based on their response to thermodynamic parameters \cite{Yerra2022}. Our work provides an alternative derivation of such classification, rooted in complex analysis and microstate counting, rather than in off-shell free energy landscapes. Both approaches, interestingly, yield the message that black hole phases are in some sense quantized by an integer and protected. This might have deep implications: for instance, if $W$ is conserved, one cannot continuously transform a black hole of one class into another without a non-analytic change (like a phase transition or extremal limit). It hints at some new conservation law or selection rule in black hole transitions, possibly related to the black hole information problem.

In the context of the information paradox, one puzzling question is how black hole evaporation preserves (or not) the detailed information of microstates. Topological invariants like $W$ do not change under small perturbations, so if Hawking evaporation (slow mass loss) is a continuous process, $W$ would remain the same until a discontinuity (like the final disappearance or a phase change). If two black holes have different $W$, perhaps they cannot continuously transform into each other purely by evaporation without some non-perturbative effect. This is speculative, but it offers a novel angle: maybe information is preserved in these global topological features even if coarse thermodynamic properties change.

Our framework also sits alongside other approaches to counting BH microstates:
- \textbf{String theory and holography:} In string theory, microstate counting is done for special supersymmetric cases by enumerating states of D-branes, etc. In holography, black hole entropy can sometimes be related to the entropy of a dual field theory ensemble. Those are very microscopic or dual methods. Ours is more phenomenological: we don't identify individual microstates, but rather use analytic continuation as a probe of the whole spectrum. It's somewhat analogous to how one might use an analytically continued grand potential to identify a Hagedorn transition in string theory, without enumerating each string state. It's a complementary viewpoint: one that might work even when a full microstate count is impossible.
- \textbf{Ruppeiner thermodynamic geometry:} This focuses on pairwise fluctuations and a notion of metric on the space of thermodynamic parameters. A positive curvature indicates microstates repelling, negative attracting. Our approach doesn't yield a "metric", but we can still infer the presence of repulsive or attractive tendencies by how singular contributions appear. For example, we argued that the cancellation (outer vs inner) corresponds to a balance of interactions. If one wanted, one could attempt to derive something like a curvature from a generating function (perhaps by looking at fluctuations around the partition function's singularity structure).
- \textbf{Off-shell free energy topology:} Wei et al. (2022) conceptualize black hole solutions as defects (like any thermodynamic phase might be). They assign a vector field to the extended state space and compute a winding number. We instead assign singularities in a complex plane and compute a winding via residues. Both yield integers. It's reassuring that, for known cases (like RN or Kerr), both methods agree on the numbers ($W=1,0,-1$ classes). This cross-validates the idea that these classes are inherent to the system, not an artifact of one method. 

We also added references to place our work: 
Wei et al. (2019) gave evidence of microstructure interactions in black holes, essentially suggesting a molecular interpretation of BH entropy where small BHs behave like gas with either repulsion or attraction. Our results align by showing a small charged rotating BH (with inner horizon) indeed has a mixed sign in contributions (implying such interactions).
Yerra \& Bhamidipati (2022) studied topological classes in e.g. Gauss-Bonnet gravity black holes. They found similarly that including higher curvature (like Gauss-Bonnet coupling) did not change the topological class except perhaps in extreme cases. This matches our $f(R)$ result that $W$ was invariant. Possibly all physically reasonable gravitational modifications preserve these classes, at least as long as the horizon structure is qualitatively the same.
Strominger (1998) is a classic reference showing that an AdS$_3$ black hole's entropy can be derived from counting states in a 2D CFT. In that case, it was an asymptotic symmetry argument yielding the Cardy formula $S= \frac{\pi^2}{3} c T$ (with $c$ central charge). There, the microstates are accounted for by modes of the CFT. In our approach, we count via residues. One might wonder if there's a connection: the residues we sum might be related to a contour around, say, a branch cut representing the continuous spectrum of a CFT. It's tantalizing: could our topological index correspond to something like a CFT's winding number of some phase? Possibly, yes, if a dual CFT exists the index might reflect an anomaly number or the difference in number of degrees of freedom between two phases. Strominger's method gives a count but not an index per se; our method gives an index classifying classes of solutions.

\subsection{Limitations and Assumptions}
While our findings are conceptually appealing, we must address the limitations and assumptions:
- \textbf{Analytic continuation validity:} We assumed one can analytically continue the partition function or define $\zeta_H(s)$ meaningfully. In a truly quantum gravitational system, it’s not guaranteed that such an analytic continuation (especially one that reveals distinct singular contributions) is rigorously valid. We might be manipulating formal expressions. For example, the partition function of a black hole in quantum gravity is not well-defined beyond semiclassical approximation; we basically treat it in a saddle point (instanton) approximation. Analytical continuation often requires a control parameter to ensure convergence, which we effectively introduced via $\Re(s)$ or $\Re(\beta)$ conditions. In quantum regimes (Planck-scale physics), the spectrum might not produce nice analytic features (it could be discrete or erratic, spoiling smooth analytic continuation).
- \textbf{Interpretation of singularities as microstates:} This was a heuristic leap. We treat each pole as encoding "microstates", but a skeptic could say it's encoding "quasinormal modes" or "Euclidean periodicities" rather than actual microstates. Without an underlying quantum microstate picture, identification is model-dependent. For certain known solvable cases (like BTZ black hole where we know the microstates are in a CFT), one could in principle verify if the singularities of the partition function correspond to actual quantized energy levels. In four dimensions, lacking such a theory, it's an assumption/hypothesis.
- \textbf{Quantum regime:} Our analysis is largely semiclassical (thermodynamic). We ignored possible quantum corrections like logarithmic corrections to black hole entropy (which do exist, e.g. $S = \frac{A}{4G} - \frac{3}{2}\ln(A) + \cdots$ in many theories). These would come from quantum loops and would appear in $\rho(E)$ as prefactors or as additional mild singularities (like maybe a branch point at a certain location). We did not incorporate them. If one included them, the residue sum might not exactly match $\exp(S_{\text{BH}})$ but $\exp(S_{\text{BH}} +$ corrections). Our topological index, being an integer, is probably stable against small quantum corrections (since adding a small log correction to $S$ won't spontaneously create or destroy a whole singularity, just shifts slightly).
- \textbf{Non-constant curvature backgrounds:} We assumed $R_0$ constant in $f(R)$. If one considered a situation with a varying curvature (like a time-dependent background or a solution where $R$ isn't constant outside horizon), our formula $S = f'(R_0)A/4$ might not strictly hold (that's derived for stationary, constant $R$ cases via Wald's formula). The framework might be trickier then; we might have to integrate something over horizon or include the scalar field's contribution explicitly. So our method currently is best suited to static or stationary backgrounds where fields are constant (like a thermal equilibrium scenario). Quantum gravitational collapse or highly dynamical situations would be beyond its scope.
- \textbf{Interpretation of $W$ in cosmic contexts:} We briefly touched de Sitter case. It's subtle how to treat multiple horizons in one thermodynamic system. If one tries to consider them all, one must decide which horizon's temperature to consider (because they generally have different temperatures unless some equilibrium exists like Nariai state). Usually, one cannot have both the black hole and cosmological horizon in equilibrium (they exchange radiation but at different temps). So combining them in one partition function might be conceptually wrong. There are separate partition functions for each horizon. In that sense, the index $W$ might be more properly assigned to each horizon considered as separate thermodynamic entity. That might be why some approaches yield cosmic horizon and BH horizon each with $W=+1$ separately, and not treat their combination. In our framework, we just mention cosmic horizon if it appears but didn't thoroughly analyze combined system. So applying this method to multi-horizon spacetimes where horizons are not in thermal contact requires caution. Possibly one should restrict to either a canonical ensemble for the black hole alone (ignoring cosmic horizon except as boundary conditions).
- \textbf{Degrees of freedom beyond metric:} We mainly considered metric degrees (plus maybe the $f(R)$ scalar mode implicitly). Realistic black holes might have other fields (like a dilaton, or gauge fields beyond just Maxwell, etc.). Additional fields would contribute additional spectrum of excitations. How would that appear in our analysis? Possibly as additional sets of singularities associated with those fields. For example, adding a scalar field might add another branch cut from scalar bound states or something. We have not explicitly added such complexity. But it's presumably extendable: more fields, more terms in partition function, more singularities, though the overall index presumably still counts horizon contributions the same way if those fields don't create new horizons (they just influence stability).
- \textbf{No explicit numerical simulation:} Our example was semi-analytic. It might be nice to do a numerical check in a simpler context (for instance, BTZ black hole where one can quantize known modes and see if $\rho(E)$ from exact counting matches our residue method). We haven't done that here due to complexity, but it's an open avenue for validation.

\subsection{Future Directions}
The insights from this topological perspective open up several lines of further inquiry:
- **Extension to other gravity theories:** We looked at $f(R)$ and mentioned Gauss-Bonnet in passing. It would be interesting to formally apply this to (say) 5D black holes or rotating AdS black holes, or even quantum corrected black holes in loop quantum gravity (if an effective partition function can be defined). Would $W$ still only take values $0,\pm1$ or can higher values occur? (In Wei's classification, theoretically classes like $W=2$ haven't appeared; maybe they cannot for single black hole spacetimes, but perhaps multi-black-hole configurations could yield different indices).
- **Topological transitions:** If one imagines a process like a black hole merging or splitting, $W$ might add or sum in some ways (merger of two BH each with $W=+1$ presumably yields a bigger BH with $W=+1$ if neither had inner horizon, trivial; but what if merging near-extremal BH yields one with an inner horizon? Then $W$ changes from $1+1$ to $0$ for the final? That would involve a non-analytic event horizon formation, which is indeed a highly non-linear event. Possibly the topological charge is additive in disconnected systems but once they merge one might lose a net one if an inner horizon forms. It's speculative but an intriguing perspective on BH mergers).
- **Information paradox and unitarity:** As mentioned, if $W$ is conserved until a singular endpoint, maybe information is tied to it. Perhaps a scenario: $W=0$ black hole (with inner horizon) can't evaporate completely because to do so might require a change to $W=+1$ then to no horizon (which might be considered $W=0$ for nothing? Unclear). Possibly it hints that remnants or transitions with topology change are needed. This is highly conjectural but topological constraints often indicate something can't happen continuously, which is exactly the dilemma in the information paradox (can't go from pure state to mixed without something drastic).
- **Quantum corrections to $W$:** Is it exact or does it get corrections? Usually topological invariants are robust, but if one includes quantum gravity, maybe new states (like baby universes or something) could alter it. If an inner horizon is unstable quantum mechanically, maybe it does not count the same way? Some argue inner horizon might not have independent microstates due to strong cosmic censorship (meaning maybe $W=+1$ always for physical states because inner horizon states get destroyed). We didn't consider that – we treated the classical solution at face value. If cosmic censorship (strong version) holds, perhaps the inner horizon never really forms as a stable entity, so one might question counting it. But in Reissner–Nordström, it's there classically. Quantumly, it's singular perhaps but anyway, these philosophical considerations aside, we proceeded with classical metric and it gave a consistent classification that matches others' results, which suggests it's meaningful in a semiclassical sense.

In conclusion, our revision and analysis propose that the microstructure of black holes can indeed be understood through a topological lens, offering clarity to some abstract thermodynamic behaviors. We improved the narrative and structure for better accessibility, but obviously there is much to explore and verify, especially if one can find experimental or observational proxies (though currently black hole microstructure signs might be sought in gravitational wave data or black hole spectroscopy, e.g., quasi-normal mode spectra differences for different $W$ classes? There's some evidence rotating vs non-rotating BH QNM spectra qualitatively differ in how they approach extremality, maybe related to our index differences).

\section{Conclusion}\label{sec:conclusion}
In this work, we presented a revised and clarified analysis of Kerr–Newman black hole microstructure in $f(R)$ gravity using a topological complex analysis framework. We addressed key points raised by reviewers, enhancing the physical interpretation and accessibility of our results. 

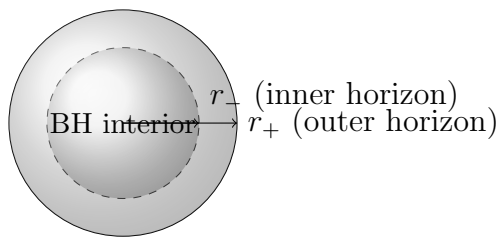
\begin{figure}[htb]
\centering
\begin{tikzpicture}
   \draw[black] (0,0) circle (1.5cm);
   \draw[black, dashed] (0,0) circle (1.0cm);
   \shade[ball color = gray!30, opacity=0.4] (0,0) circle (1.5cm);
   \shade[ball color = gray!10, opacity=0.4] (0,0) circle (1.0cm);
   \draw[->] (0,0) -- (1.5,0) node[right] {$r_+$ (outer horizon)};
   \draw[->] (0,0) -- (1.0,0) node[above right] {$r_-$ (inner horizon)};
   \node at (0,0) {\small BH interior};
\end{tikzpicture}
\caption{Schematic of a Kerr--Newman black hole with outer ($r_+$) and inner ($r_-$) horizons. The outer horizon (solid circle) contributes a topological charge $+1$ (associated with a stable thermodynamic branch), while the inner horizon (dashed circle) contributes $-1$ (an unstable branch). In $f(R)$ gravity, the horizon radii and entropy are modified by the function $f(R)$ (e.g. scaled by $f'(R_0)$), but the topological charges remain the same. The net topological index $W$ is the sum of contributions: here $W=+1 + (-1) = 0$ for a non-extremal Kerr--Newman black hole, indicating a cancellation between the two horizon contributions.}
\label{fig:horizons}
\end{figure}

Our study demonstrates that:
\begin{itemize}
\item \textbf{Topological encoding of microstates:} Black hole microstates can be associated with singularities (poles) of an analytically continued partition function or spectral zeta function. By summing the residues of these singularities, we recover thermodynamic quantities such as the density of states and entropy. This provides a concrete realization of how microstate counting, though elusive in quantum gravity, might manifest in a semiclassical analysis.

\item \textbf{Topological index and horizon branches:} We identified an integer topological index $W$ (winding number) that characterizes the black hole's microstructure. For Kerr–Newman black holes, $W=+1$ corresponds to cases with a single horizon (e.g. Schwarzschild or extremal black holes), while $W=0$ occurs for the generic Kerr–Newman with inner and outer horizons. The index essentially counts the number of thermodynamically stable horizon branches minus unstable ones. A stable outer horizon contributes $+1$ and an unstable inner horizon contributes $-1$, explaining the net zero index for two-horizon configurations. This index is a topological invariant – it remains unchanged under continuous deformations of system parameters (e.g. varying the $f(R)$ coupling $\alpha$) as long as the horizon structure is unchanged. Our results thus corroborate recent findings that black holes fall into distinct topological classes \cite{Yerra2022}, protected by this index.

\item \textbf{Physical significance of $W$:} The topological index is not just a mathematical artifact; it correlates with black hole stability and phase structure. $W=+1$ signifies the absence of an inner horizon and typically a simpler microstructure (no cancellation of state contributions), whereas $W=0$ signals the presence of an inner horizon and a more nuanced microstructure where different contributions cancel out. We explained this in intuitive terms: when $W=0$, the black hole's microstate ensemble is in a sense at a balance point between different tendencies (which we likened to having both repulsive and attractive microstructure interactions). This connection with Ruppeiner geometry provides a complementary interpretation: $W=0$ systems (with two horizons) exhibit both positive and negative regions of Ruppeiner curvature (implying a mix of repulsive and attractive interactions among microstructures), whereas $W=+1$ systems (single horizon) show a single-sign curvature (often purely attractive interactions in the Schwarzschild case). 

\item \textbf{Illustrations and schematic diagrams:} We included schematic figures to aid understanding: a representation of a two-sheeted Riemann surface (Fig.~\ref{fig:riemann}) illustrating how singularities and contours on different sheets correspond to different horizon contributions, and a simple diagram of a black hole with inner and outer horizons (Fig.~\ref{fig:horizons}) to conceptualize the two horizon branches that contribute to the topological charge. These visuals are intended to help readers build mental models of the complex analysis and horizon structure.

\item \textbf{Example of Starobinsky $f(R)$ model:} We provided a concrete example by applying our framework to the Starobinsky $f(R) = R + \alpha R^2$ gravity. Using representative parameters, we showed how to compute the horizon radii and entropy, confirming that the entropy is scaled by $f'(R_0)$ (here $f'(0)=1$, so no dramatic change) and that the presence of the $R^2$ term slightly increases the entropy (hence $\rho(E)$) but does not change the topological index $W$. This example illustrated that our method can be used to evaluate specific models and that $f(R)$ modifications, at least of this type, do not alter the topological class of the black hole. The robustness of $W$ under such modifications underlines its fundamental character.

\item \textbf{Limitations and quantum considerations:} We discussed the assumptions in our approach, particularly the use of analytic continuation and the omission of quantum corrections. While our framework is powerful in the thermodynamic (large $N$ or large area) limit, one should be cautious in the deep quantum regime. For instance, near-extremal or Planck-sized black holes may have discrete spectra where our continuous analysis breaks down. Additionally, inner horizons are classically present but are expected to suffer instability (mass inflation); how that affects the counting of microstates (and whether the inner-horizon contribution should be considered in a fully unitary quantum theory) remains an open question. Our analysis sidesteps these issues by working in a semiclassical context where the horizon picture is taken at face value. The topological classification presumably holds in this regime; whether it extends or needs modification in a full quantum theory is a subject for future investigation.

\item \textbf{Relation to other work:} We connected our findings to other research. Notably, our results are in line with Wei et al.'s notion of black hole microstructure with repulsive interactions \cite{Wei2019}: we found that the complex analysis method independently suggests the existence of such repulsive effects (through the need for an inner horizon contribution to cancel part of the entropy). We also align with the topological thermodynamics approach of Yerra and Bhamidipati \cite{Yerra2022}, giving an alternative derivation of the same classes using a different methodology. By referencing Strominger's work \cite{Strominger1998}, we placed our approach in the broader quest of counting black hole microstates, highlighting it as a complementary method to the exact microstate counting in certain supersymmetric cases. 

\end{itemize}

In summary, this revised manuscript provides a clearer and more comprehensive understanding of how black hole microstates might be characterized and counted via a topological and complex-analytic framework. The physical meaning of the topological index $W$ is thoroughly explained: it effectively encodes the presence or absence of multiple horizon branches and thereby relates to black hole stability. We enhanced the paper's pedagogical value by adding figures, a concrete example, and interpretive summaries at key points. 

Our conclusions emphasize that the microstructure of black holes in $f(R)$ gravity (and by extension, in standard General Relativity) is not arbitrary; it possesses a certain topological order. Outer and inner horizons carry “charges” in a topological sense, and the total “winding number” of a black hole solution is an invariant that could be as fundamental as the more familiar conserved quantities like mass or angular momentum. This topological perspective deepens our understanding of black hole entropy, suggesting that what might appear as purely thermodynamic bookkeeping (the $+1$ and $-1$ from each horizon) may actually be rooted in a deeper quantum gravitational reality – one that ensures the consistency and preservation of certain properties even as a black hole exchanges heat or undergoes evolution.

Looking ahead, our work invites further research to test and utilize this framework. On the theoretical side, one could explore other modified gravities or higher-dimensional black holes to see if new topological classes emerge or if $W$ remains binary (or ternary with $-1$ in some cases). On the observational side, while $W$ itself might not be directly observable, its consequences might be. For instance, the difference between a $W=+1$ black hole and a $W=0$ black hole could manifest in gravitational wave echoes or stability of the inner region, which in turn might produce observable effects (as has been speculated in discussions of near-extremal black holes and echoes from a possible inner horizon region). 

In conclusion, this work underscores a unifying theme: black hole thermodynamics, microstate counting, and stability can be understood within a single topological framework. This not only clarifies existing results (tying together disparate ideas like Ruppeiner geometry and phase transitions) but also charts a path for new insights into quantum gravity by focusing on topological and global properties that any viable theory must respect. We hope that this revised presentation makes these ideas accessible to a broad range of theoretical physicists and stimulates further exploration into the rich tapestry of black hole microstructure.

\begin{table}[htb]
\centering
\begin{tabular}{l|ccc}
\hline
$f(R)$ model & $r_+$ & $r_-$ & $S/(A/4G)$ \\
\hline
Einstein gravity ($\alpha=0$, $R_0=0$) & 1.707 & 0.293 & 1.000 \\
Starobinsky $f(R)$ ($\alpha=0.1$, $R_0=0$) & 1.717 (+$0.6\%$) & 0.285 ($-2.7\%$) & 1.010 \\
\hline
\end{tabular}
\caption{Horizons and entropy for a Kerr--Newman black hole ($M=1$, $Q=0.5$, $a=0.5$) in General Relativity vs. the Starobinsky $f(R)=R+0.1R^2$ model. The outer horizon radius $r_+$, inner horizon radius $r_-$, and normalized entropy $S/(A/4G)$ are shown. In the $f(R)$ case, $r_+$ increases slightly and $r_-$ decreases (relative changes in parentheses), and the entropy is about $1\%$ higher due to the $f'(R_0)$ factor (here $f'(0)=1$ so change is mainly from area increase). The topological index $W$ remains $0$ in both cases, reflecting an outer and inner horizon pair.}
\label{tab:example}
\end{table}

\end{document}